\documentclass[twocolumn,twoside,slac_two]{revtex4}
\usepackage{graphicx}
\usepackage{fancyhdr}
\begin{document}

\title{Searches for, and Properties of, New Charmonium-like states}


\author{Paul~D.~Jackson\footnote{Paul.Jackson@roma1.infn.it}}
\affiliation{Universita di Roma ``La Sapienza'' and INFN Rome \\
Dipartimento di Fisica, Piazzale Aldo Moro, 2, Roma, I-00185, Italy
}

\begin{abstract}
We review the recent $B$-factory measurements of new states which, 
in some cases, exhibit Charmonium-like properties, and in other cases suggest
the existence of a new spectroscopy. Several theoretical interpretations of
the new states have come to the fore although, at time of writing, we
are no closer to untangling the nature of most of the particles making
up the observed new zoo of states.
 
\end{abstract}

\maketitle

\thispagestyle{fancy}


\section{Introduction}

Although the Standard Model of elementary particles is well 
established, strong interactions are not yet fully under control. We 
believe QCD is the field theory capable of describing them, but we are 
not yet capable, in most cases, to make exact
predictions. Systems that include heavy quark-antiquark pairs 
(quarkonia) are an ideal and unique laboratory to probe both the high 
energy regimes of QCD, where an expansion in terms of the coupling 
constant is possible, and the low energy regimes, where 
non-perturbative effects dominate.

Recently, this field has experienced a rapid expansion with a 
wealth of new data coming in from diverse sources:
data on quarkonium formation from dedicated experiments (BES at 
BEPC, KEDR at VEPP-4M CLEO-c at CESR), clear samples produced by high 
luminosity $B$-factories (PEP-II and KEKB), and very large samples produced 
from  gluon-gluon fusion in $\rm{p\bar{p}}$ annihilations at the Tevatron 
(CDF and D0 experiments).

\begin{figure}[ht]
%\centering
\includegraphics[width=72mm]{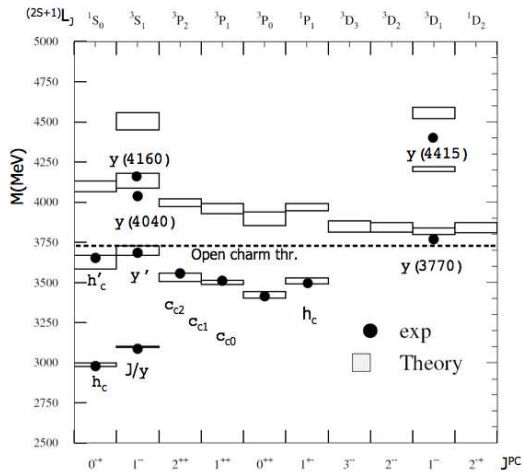}
\caption{Charmonium states with
$L<=2$. The theory predictions are according to the potential models
described in Ref.~\cite{Brambilla:2004wf}.
    \label{fig:charmon}}
\end{figure}

%In this review I will first summarize recent developments in the 
%understanding of heavy quarkonium states which have a well established 
%quark content.
%\begin{figure}[ht]
%\centering
%\includegraphics[width=80mm]{bottom_proceed.eps}
%\caption{Bottomonium states with
%$L<=2$. The theory predictions are according to the potential models
%described in Ref.~\cite{Brambilla:2004wf}.
%    \label{fig:bottom}}
%\end{figure}

This paper will be devoted to reviewing the experimental 
evidence of new states that might be aggregations of more than just a 
quark-antiquark pair. Although the possibility of bound states of 
two quarks and two antiquarks or of quark-antiquark pairs and gluons 
has been predicted since the very start of the quark 
model~\cite{GellMann:1964nj}, 
no observed state has yet been attributed to one of them:  
such an achievement would be a major step in the understanding of 
the strong interaction.
\begin{table*}[!htb]
\begin{center} 
\caption{ 
Most recent determination of the  $J^{PC}=1^{--}$  charmonium states from 
BES~\cite{Ablikim:2007gd}, compared to the 2006 edition of the PDG~\cite{Yao:2006px} 
}
\label{tab:rscan} 

\begin{tabular}{|c|c|c|c|c|c|} \hline \hline
        &         &$\psi(3770)$  &$\psi(4040)$&$\psi(4160)$&$\psi(4415)$\\ \hline
$M$     &PDG2006  &3771.1$\pm$2.4&4039$\pm$1.0&4153$\pm$3  & 4421$\pm$4 \\
(MeV/$c^2$)          &BES '07 &3771.4$\pm$1.8& 4038.5$\pm$4.6 & 4191.6$\pm$6.0 & 4415.2$\pm$7.5\\\hline\hline
$\Gamma_{tot}$&PDG2006  &23.0$\pm$2.7&80$\pm$10    &103$\pm$8    &     62$\pm$20   \\
(MeV)         &BES '07      &25.4$\pm$6.5&81.2$\pm$14.4&72.7$\pm$15.1&73.3$\pm$21.2\\\hline\hline
\end{tabular} 
\end{center}
\end{table*}

Currently the most credible explanations for the possible states, beyond the mesons and the 
baryons, are (find a full review in~\cite{Brambilla:2004wf}):
\begin{itemize}
\item hybrids: bound states of a quark-antiquark pair and a number of 
gluons. The lowest lying state is expected to have quantum numbers 
$J^{PC}=0^{+-}$. The impossibility of a quarkonium state to assume 
these quantum numbers (see below) makes this a unique signature for 
hybrids. Alternatively, a good signature would be the preference to 
decay into a quarkonium and a state that can be produced by the excited 
gluons (e.g. $\pi^+\pi^-$ pairs).
\item molecules: bound states of two mesons, usually represented as 
$[Q\bar{q}][q^{\prime}\bar{Q}]$, where $Q$ is the heavy quark. The 
system would be stable if the binding energy would set the mass of the 
states below the sum of the two meson masses.
While this could be the case for when $Q=b$, this does not apply for 
$Q=c$, where most of the current experimental data are. In this case 
the two mesons can be bound by pion  exchange. This means that only 
states decaying strongly into pions can bind with other mesons (e.g. 
there could be $D^*D$ states), and 
that the bound state could decay into it's constituents.
\item tetraquarks: a quark pair bound with an antiquark 
pair, usually represented as $[Qq][\bar{q^{\prime}}\bar{Q}]$. A full 
nonet of states is predicted for each spin-parity, i.e. a large number 
of states are expected. There is no need for these states to be close to 
any threshold.
 \end{itemize}
In pursuing a further understanding of these states one must also beware of threshold effects, 
where amplitudes might be enhanced when new hadronic final states become 
possible.

Herein, we summarize the latest findings on the 
spectroscopy of the known heavy quarkonium states and 
the state of the art of the understanding of other states which 
might not fit in the ordinary spectroscopy.

\section{Heavy quarkonium spectroscopy}
The heavy quark inside these bound states has low 
enough energy that the corresponding
spectroscopy is close to the non-relativistic interpretations of the 
atoms. The quantum numbers that are more appropriate to characterize a 
state are therefore, in decreasing order of energy splitting among 
different eigenstates, the radial excitation ($n$), the spatial angular 
momentum $L$, the spin $S$ and the total angular momentum $J$. Given 
this set of quantum numbers, the parity and charge conjugation of the 
states are derived by $P=(-1)^{(L+1)}$ and $C=(-1)^{(L+S)}$. 
Figure~\ref{fig:charmon} shows the mass and 
quantum number assignments of the well established charmonium states.

\subsection{Charmonium spectroscopy}
Figure~\ref{fig:charmon} shows that all the predicted states below 
open charm
 threshold have been observed, leaving the search open only to states above the
 threshold. In this field the latest developments concern the measurement of 
the paramaters and the quantum number assignment for the $J^{PC}=1^{--}$ 
states.

The BES collaboration has recently performed a fit to the $R$ scan 
results which takes into account interference between resonances more 
accurately~\cite{Ablikim:2007gd}. The updated parameters are reported in 
Tab.~\ref{tab:rscan}, compared with the most recent determinations.

The $J^{PC}=1^{--}$ assignment does not unambiguously identify the state, 
since both $^{2S+1}L_J$=$^3D_1$ and $^3D_1$ states would match it. The 
recent observation from Belle of the first exclusive decay of the 
$\psi(4415)\to DD^*_{2}(2460)$~\cite{belle:2007fq}, shows that this 
meson is predominantly $D$ wave. At the same time the study from CLEO-c 
of the $\psi(3770)\to\chi_{cJ}\gamma$~\cite{Briere:2006ff} confirms the dominance of the $D$ wave also in this meson. 
Both these assignments confirm the theoretical predictions as shown in Fig.~\ref{fig:charmon}.

\begin{figure}[ht]
\centering
\includegraphics[width=80mm]{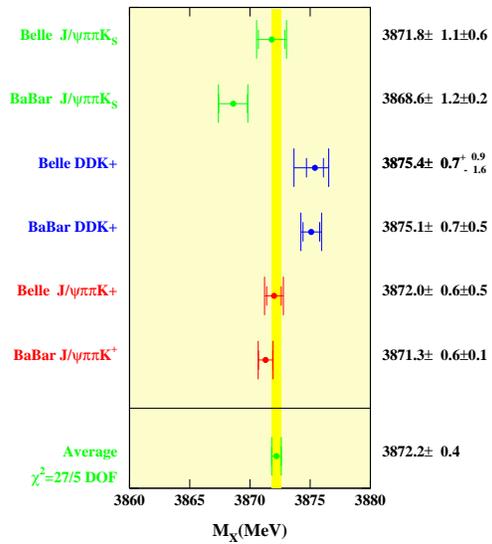}
 \caption{\it Measured mass of the $X(3872)$ particle. The different 
production modes ($B^0\to XK_S$ and $B^-\to X K^-$)
and the different decay modes ($X\to J/\psi\pi\pi$ and $X\to 
D^{*0}D^0$) are separated.} 
\label{fig:xmass}
\end{figure}

\section{Non-standard charmonium states}
\subsection{The {\boldmath $X(3872)$}}

The $X(3872)$ was the first state that was found not to easily fit charmonium spectroscopy.
 It was initially observed decaying into $J/\psi\pi^+\pi^-$ with a mass just beyond the open charm 
threshold~\cite{Choi:2003ue}. The $\pi^+\pi^-$ invariant mass distribution 
preferred the hypothesis of a 
$X(3872)\to J/\psi\rho$ decay, which would have indicated that if this 
were a charmonium state, the decay would 
have violated isospin. Since it would be quite unusual to have the dominant decay to be isospin violating,
a search of the isospin partner $X^+\to J/\psi\rho$ was conducted 
in vain by BaBar~\cite{Aubert:2004zr}.
In the meanwhile the decay $X\to J/\psi\gamma$ was 
observed~\cite{Aubert:2006aj}, implying
positive intrinsic charge conjugation.

The most recent developments concern the final assessment of the $J^{PC}$ of this particle and the indication
that the $X(3872)$ is actually a doublet. The CDF collaboration has performed the full angular analysis
of the $X\to J/\psi\pi\pi$ decay~\cite{Abulencia:2006ma} concluding 
that $J^{PC}=1^{++}$ and $2^{-+}$ are 
the only assignments consistent with data. They also confirmed that the decay proceeds through a $\rho$ intermediate state.

\begin{table*}[htb]
\begin{center} 
\caption{ 
Measured $X(3872)$ branching fractions, separated by production and decay mechanism. The ratio of the 
measurements in the two production mechanisms are also reported as $R_{0/+}=BF(B\to XK^-)/BF(B\to XK^0)$. 
}
\label{tab:xsec} 
\begin{tabular}{|l|c|c|} \hline 
                & BaBar & Belle \\ \hline
BF($B\to XK^-$)BF($X\to J/\psi\pi\pi$)$\times 10^5$&  0.84$\pm0.15\pm0.07$~\cite{ref:newX3872}  & 1.05$\pm$0.18~\cite{Choi:2003ue}\\ 
BF($B\to XK^0$)BF($X\to J/\psi\pi\pi$)$\times 10^5$&  0.35$\pm0.19\pm0.04$~\cite{ref:newX3872}  &$0.99\pm0.33$ \\   
BF($B\to XK^-$)BF($X\to D^{*0}D^0$)$\times 10^5$   &  $17\pm4\pm5$~\cite{babar:2007rv}   &10.7$\pm 3.1^{+1.9}_{-3.3}$~\cite{Gokhroo:2006bt} \\   
BF($B\to XK^0$)BF($X\to D^{*0}D^0$)$\times 10^5$   &  $22\pm10\pm4$~\cite{babar:2007rv}  &17$\pm7^{+3}_{-5}$~\cite{Gokhroo:2006bt}\\  
\hline
$R_{0/+}$ with $X\to J/\psi\pi\pi$              & $0.41\pm0.25$~\cite{ref:newX3872}& $0.94\pm0.26$ \\ 
$R_{0/+}$ with $X\to D^{*0}D^0$                 & $1.4\pm0.6$~\cite{babar:2007rv}  & -- \\ \hline
\end{tabular} 
\end{center}
\end{table*}

 As far as the mass and width of the $X(3872)$ are concerned, BaBar has 
published an analysis of the $B\to XK$
 decays with $X\to D^{*0}D^0$~\cite{babar:2007rv} while Belle has 
updated the mass measurements in $X\to J/\psi\pi\pi$
 decays~\cite{bellex}. The summary of all available mass measurements 
is shown in Fig.~\ref{fig:xmass} where 
 the measurements are separated by production and decay channel. There 
is an indication that the particle
 decaying into $J/\psi\pi\pi$ is different from the one decaying into 
$D^{*0}D^0$, their masses differing
by about 4.5 standard deviations.
 
%%***UPDATE BaBar X(3872) analysis***//
Recently, BaBar published an updated analysis of the $X(3872)$ with their complete dataset~\cite{ref:newX3872}, studying
the discovery mode of the $X(3872)\rightarrow J/\psi\pi^+\pi^-$. Updated branching fraction results (see Table~\ref{tab:xsec} for details)
and updated mesaurements of $\Delta m$ ($=2.7\pm1.6\pm0.4$ Mev/c$^2$) and the natural width ($\Gamma<3.3$ MeV/c$^2$) were provided. 
The central value for the $X(3872)$ natural width was found to be
$\Gamma=$($1.1 \pm1.5 \pm 0.2$)MeV. 
All $X(3872)$ branching fractions measurements, in $J/\psi\pi\pi$ and $D^{*0}D^0$, are summarized in Tab.~\ref{tab:xsec}.

\subsection{The {\boldmath $1^{--}$} family}
The easiest way to assign a value for $J^{PC}$ to a particle, is to observe its production via $e^+e^-$ 
annihilation, where the quantum numbers must be the same as the radiated photon:  $J^{PC}=1^{--}$. $B$ factories
can investigate a large range of masses for such particles by looking for events where the initial state 
radiation brings the $e^+e^-$ center-of-mass energy down to the particle's mass (the so-called 'ISR' events). 
Alternatively, 
dedicated $e^+e^-$ machines, like CESR and BEPC can scan directly the center-of-mass energies of interest.

\begin{figure}[ht]
\centering
\includegraphics[width=80mm]{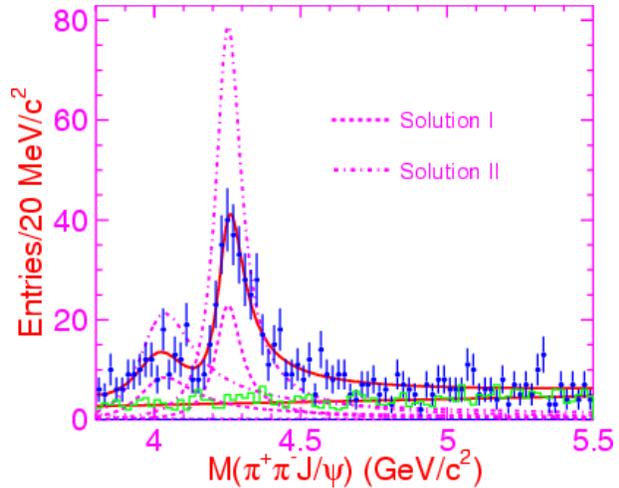}
 \caption{\it $J/\psi\pi^+\pi^-$  invariant mass in ISR production.}
\label{fig:belle1mm}
\end{figure}

\begin{figure}[ht]
\includegraphics[width=80mm]{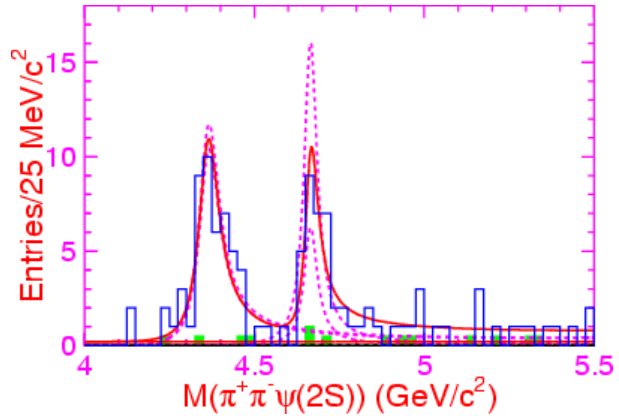}
 \caption{\it  $\psi(2S)\pi^+\pi^-$ invariant 
mass in ISR production.}
\label{fig:belle1mm2S}
\end{figure}

The observation of new states in these processes started with the discovery of the $Y(4260)\to 
J/\psi\pi^+\pi^-$ 
by BaBar~\cite{Aubert:2005rm}, promptly confirmed both in the same production process~\cite{He:2006kg} and in 
direct production by CLEO-c~\cite{Coan:2006rv}. The latter paper also reported evidence for 
$Y(4260)\to J/\psi\pi^0\pi^0$ and some events of $Y(4260)\to J/\psi K^+K^-$.

While investigating whether the $Y(4260)$ decayed to 
$\psi(2S)\pi^+\pi^-$ BaBar did not find evidence for such a decay. Instead, they
discovered a new $1^{--}$ state, the $Y(4350)$~\cite{Aubert:2006ge}. 
While the absence of $Y(4260)\to \psi(2S)\pi^+\pi^-$ decays could be 
explained if the pion pair in the $J/\psi\pi^+\pi^-$ decay were 
produced 
with an intermediate state that is too massive to be produced with a 
$\psi(2S)$ (e.g. an $f^0$), the absence of $Y(4350)\to 
J/\psi\pi^+\pi^-$ is still to be understood, more statistics might be 
needed in case the $Y(4260)$ decay hides the $Y(4350)$.

Recently, Belle has published the confirmation of all of these $1^{--}$ 
states~\cite{belle:2007sj,belle:2007ea}. Furthermore, they have
unveiled a new state that was not clearly visible in  the BaBar data due to the 
limited statistics: the $Y(4660)$. Figures~\ref{fig:belle1mm} 
and~\ref{fig:belle1mm2S} show the 
published invariant mass spectra for both the $J/\psi\pi^+\pi^-$ and 
the $\psi(2S)\pi^+\pi^-$ decays. 
\begin{figure*}[t]
\includegraphics[width=155mm]{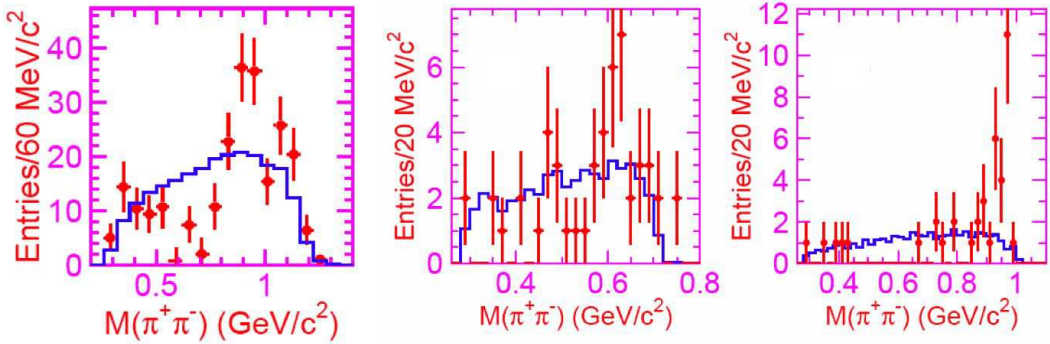}
 \caption{\it Di-pion invariant mass distribution in $Y(4260)\to 
J/\psi \pi^+\pi^-$ (left), $Y(4350)\to \psi(2S) \pi^+\pi^-$ (center), 
and $Y(4660)\to \psi(2S) \pi^+\pi^-$ (right) decays. 
}
\label{fig:bellepipiinv}
\end{figure*}
A critical piece of information for unravelling the puzzle is whether the 
pion pair comes from a resonant state. Figure~\ref{fig:bellepipiinv} 
shows the di-pion invariant mass spectra published by Belle for all the 
regions where new resonances have been observed. Although the 
subtraction of the continuum is missing, there is some
indication that only the $Y(4660)$ has a well defined intermediate 
state (most likely an $f_0$), while others have a more complex 
structure.

A discriminant measurement between Charmonium states and new 
aggregation forms is the relative decay rate between these decays into 
Charmonium and the decays into two charm mesons. Searches have 
therefore been carried out for $Y\to D^{(*)}D^{(*)}$ 
decays~\cite{Abe:2006fj,collaboration:2007mb,Aubert:2007pa} without any 
evidence for a signal. The most stringent limit is~\cite{Aubert:2007pa} 
$BF(Y(4260)\to D\bar{D})/BF(Y(4260)\to J/\psi\pi^+\pi^-)<1.0 @$ 90\% 
confidence level.
\begin{figure}[ht]
\includegraphics[width=80mm]{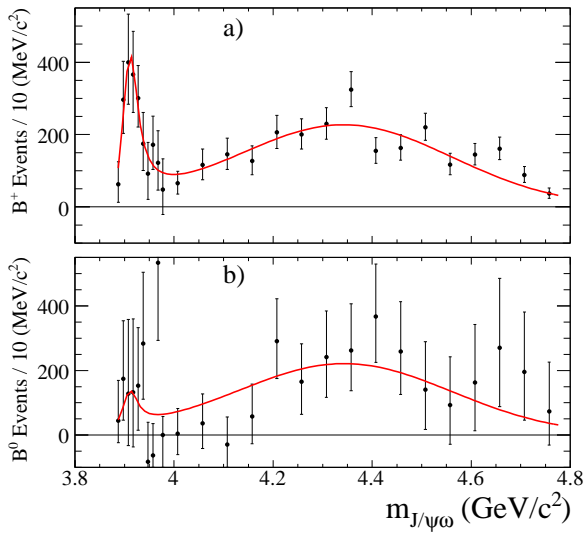}
 \caption{\it The
$J/\psi \omega$ distribution in a) $B\to J/\psi \omega K^+$ and b)  
$B\to J/\psi \omega K_S$ decays. The superimposed line is the result of 
the fit to the data.
\label{fig:babarjpsiomega}}
\end{figure}
\subsection{The {\boldmath $3940$} family}
Three different states have been observed in recent years by 
the Belle collaboration with masses close to $3940 \rm {Mev/c}^2$: one, 
named 
$X$, observed in continuum events 
(i.e. not in $Y(4S)$ decays) produced in tandem with a $J/\psi$ 
meson and decaying into $DD^{*}$~\cite{Abe:2007jn}; a second one, named 
 $Y$, observed in $B$ decays and decaying into 
$J/\psi\omega$~\cite{Abe:2004zs}; a third 
one, named $Z$ produced in two-photon reactions and decaying into 
$D$-pairs~\cite{Uehara:2005qd}. While the $X$ is consistent with both 
$J^{PC}=0^{+-}$ and $1^{++}$, the quantum number assignment of the $Y$ 
and the $Z$ states is clear: $J^{PC}=1^{++}$ and $2^{++}$ respectively.
Finally the $Y$ is the only apparently broad state 
($\Gamma=87\pm34$MeV).

Because of these quantum number assignments and their masses these 
states are good candidates for the radial excitation of the $\chi$ 
mesons, in particular the $Z(3940)$ meson could be identified with the 
$\chi_{c0}(2P)$ and the $Y(3940)$ with the $\chi_{c1}(2P)$.
The unclear points are the identification of the $X(3940)$ state and 
the explanation of why the $Y(3940)$ state does not decay preferentially 
to $D$ mesons.

The most recent development on this topic is the confirmation from 
the BaBar collaboration of the $Y(3940)\to J/\psi\omega$ 
decays~\cite{Aubert:2007vj}. The analysis utilizes the decay properties 
of the $\omega$ meson to extract a clean signal (see 
Fig.~\ref{fig:babarjpsiomega}).
The interesting part is that the mass and 
width measured in this paper are lower than when previously observed, albeit 
consistent 
($m_Y=3914.6^{+3.8}_{-3.4}(stat.)\pm1.9(sys.)$Mev$/c^2$, 
$\Gamma_Y=33^{+12}_{-8}(stat.)\pm5(sys.)$MeV), opens the interesting 
possibility that the $X$ and the $Y$ particles be the same, thus 
solving the two aforementioned open issues.

\subsection{The {\boldmath $X(4160)$}}
As we have already discussed, it is critical to investigate decay 
channels of the new states into $D$ meson pairs. Unfortunately the 
detection efficiency for $D$ mesons is low, due to the large number of 
possible decay channels. The Belle collaboration has developed a partial 
reconstruction technique 
that overcomes this limitation in the case of new states  
produced in continuum paired with known Charmonium 
states~\cite{belle:2007sy}. The Charmonium is fully reconstructed, 
while only one of the two D mesons is reconstructed. The kinematics of 
the other is inferred from the known center-of-mass energy and the 
different possibile $D$ mesons are discriminated on the basis of the 
missing mass.

This technique has allowed the confirmation of the $X(3940)$ production 
and decay, and, most interestingly, the observation of 
the $X(4160)$ state, decaying into $D D^{*}$. Given the fact that, for 
reasons yet to be understood, continuum events seem to produce  
$J^{PC}=0^{-+}$ or $1^{++}$ states in pair with the $J/\psi$ and since 
the measured mass is consistent with the expectations of a radial 
excitation of the $\eta_c$, this new state is likely to be an 
$\eta_c(3S)$.
\begin{figure}[h]
\includegraphics[width=80mm]{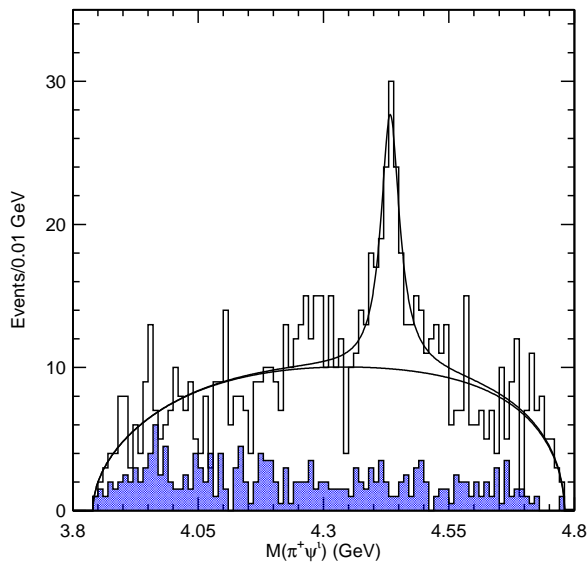}
 \caption{\it The $\psi(2S)\pi$ invariant mass distribution in 
$B\to \psi(2S)\pi K$ decays.}
\label{fig:belleZ4430}
\end{figure}
\subsection{The first charged state?: {\boldmath $Z(4430)$}}
Perhaps, the real turning point in the query for states beyond the Charmonium 
was the observation by the Belle Collaboration of a charged state
decaying into $\psi(2S)\pi^\pm$~\cite{belle:2007wg}. 
Figure~\ref{fig:belleZ4430}
shows the fit to the  $\psi(2S)\pi$ invariant mass distribution in
$B\to \psi(2S)\pi K$ decays, returning a mass $M=4433\pm4\rm{MeV/c}^2$  
and a width $\Gamma=44^{+17}_{-13}$ MeV. Due to the relevance of such 
an observation a large number of tests have been performed, breaking the 
sample in several subsamples and finding consistent results in all 
cases. Also, the possibility of a reflection of a $B\to\psi(2S)K^*$  
decay has been falsified by explicitely vetoing windows in  the $K\pi$ 
invariant mass.

In terms of quarks, such a state must 
contain a $c$ and a $\bar{c}$, but given its charge 
it must also contain at least a $u$ and a $\bar{d}$. The only open 
options are the tetraquark, the molecule or the threshold effects. The 
latter two options are possible due to the closeness of the $D_1D^*$ 
threshold. 
%\begin{figure}[h]
%\includegraphics[width=80mm]{summary.eps}
% \caption{\it Measured masses of the newly observed states, positioned 
%in the spectroscopy according 
%to their most likely quantum numbers. The charged state ($Z(4430)$) has 
%clearly no $C$ quantum number.}
%\label{fig:summary}
%\end{figure}
%
Finding the corresponding neutral state, observing a decay mode of the 
same state or at least having a confirmation of its existence, are 
critical before a complete picture can be drawn.

\section{Conclusions}
More than 30 years after its first observation, the heavy-quarkonium  is
still an exciting laboratory for understanding QCD. The study of well 
established quarkonium states 
yields information on low energy QCD while the understanding of the 
quarkonium spectroscopy, 
predictable in potential models, allows searches for different 
aggregation states than the long established mesons.

The high statistics and quality data from $B$-Factories have produced a 
very large number of new
 states whose interpretation is still a matter of debate. This paper 
attempted a categorized review of the information pertaining to these states, as we know it today. 

Many theoretical models have 
been developed to interpret 
the situation but the picture is far from complete: more precise 
predictions are needed from theory
and a systematic experimental exploration of all possibile production 
and decay mechanisms of these new states 
is still in the works. It is clear that a Super-$B$ Factory will be essential to
gain a complete understanding of the true nature of these states.
\section{Acknowledgments}
I would like to thank my colleagues in the BaBar collaboration, 
and specific members of the Belle, Cleo and CDF collaborations,
for the information I received while preparing both the talk and these proceedings.
I reserve special thanks for Professor Riccardo Faccini.

\bigskip % extra skip inserted

\end{document}